\documentstyle[12pt,PR_CDL]{article}  
\begin{document}

\def\gtsim{\hbox{$ {     \lower.40ex\hbox{$>$}
                   \atop \raise.20ex\hbox{$\sim$}
                   }     $}  }

\hfill TRI-PP-93-23

\hfill SFU-HEP-103

\hfill May 1993

\title{\bf Exploring confinement by cooling: A study of compact QED$_3$}
\vspace{0.8cm}
\author{Howard D. Trottier\\
\vspace{0.3cm}
{\em Department of Physics, Simon Fraser University,
Burnaby, B.C. Canada, V5A 1S6}\\
\vspace{0.3cm}
R.M. Woloshyn\\
{\em TRIUMF, 4004 Wesbrook Mall, Vancouver, B.C. Canada, V6T 2A3}}

\maketitle  

\setlength{\baselineskip}{5ex}

\vspace{2.0cm}
\begin{center}
{\bf Abstract}
\end{center}


The role of monopoles in the confining behavior of compact lattice
$QED_3$ is studied using an adiabatic cooling method.
Monopole-antimonopole pairs with large separation suvive
cooling and the presence or absence of such plasma monopoles
provides a useful classification of the lattice gauge field
configurations at large $\beta$. By calculating observables
in subsets of gauge field configurations which contain
or do not contain plasma monopoles it is seen that, in
compact $QED_3$, monopoles dominate the long distance physics, e.g.,
the string tension, linear confining potential
and dynamical mass generation. On the other hand, the spin-spin
interaction is essentially unaffected by monopoles.

%

\newpage

\section{INTRODUCTION}


The cooling method was first proposed as a way of exposing topological
configurations in lattice field theory\cite{berg,teper1}. More recently there
has been a lot of discussion about cooling as a probe of confinement
dynamics\cite{campo,digiac,faber,duncan90,duncan91,chu}.
This was
motivated by the finding of Campostrini \emet \cite{campo} that the string
tension, i.e., the Creutz ratio of large Wilson loops, survives a significant
amount of cooling. This invites a semiclassical interpretation for the
confinement mechanism, a theme that was carried over to a study of quantum
chromodynamics (QCD) in 2+1- dimensions by Duncan and
Mawhinney\cite{duncan90,duncan91}. However, precisely what cooling tells us
about the confinement mechanism is still not completely clear.
Teper\cite{teper2} has argued that the local nature of the cooling algorithm
inevitably leads
to persistence of long distance effects and does not necessarily reveal the
underlying dynamics.

To gain some insight into how cooling can be used to explore confinement it
seems reasonable to study a theory in which the confinement mechanism is well
understood. In this paper we focus on compact quantum electrodynamics in 2+1-
dimensions (QED$_3$). In this theory there is a semiclassical picture of
confinement\cite{polyakou,banks,duncan91}. The instantons of this
theory (which nonetheless are called monopoles due to their similarity to
magnetic monopoles in 3+1- dimensional theories) condense in the
vacuum\cite{polyakou,banks}, producing a ``magnetic" superconductor which
confines electric flux. In the lattice QED$_3$ it is found that
monopole-antimonopole pairs which are separated by more than a few lattice
spacings (unbound or plasma monopoles as we will call them) survive local
adiabatic cooling (see also Ref. \cite{duncan91}). The calculations reported
here were done at weak coupling (large $\beta$) where the density of plasma
monopole-antimonopole pairs is small and they are even absent in some of the
gauge field configurations produced by the Monte Carlo. Then by calculating
monopole properties in some detail and comparing observables in gauge field
configurations with and without plasma monopoles, one sees quite clearly the
role of monopoles in determining the long distance behavior of the theory.

Section II contains a description of the calculational methods used in this
paper. The single plaquette action is used to describe compact QED$_3$. Gauge
field configurations were generated using the heatbath algorithm. With periodic
boundary conditions it can happen that a monopole and antimonopole can
annihilate leaving ``magnetic" flux wrapped around the lattice. Once this
happens such configurations are very difficult to
change with local updating\cite{grosch} so
following Ref. \cite{dang} a global Metropolis update which injects a flux loop
is done periodically to move the system in the space of field configurations.

Cooling was carried out using an algorithm that allows only a small local
change in the gauge field at any step\cite{campo}. Monopole-antimonopole pairs
with a small separation usually annihilate quite quickly (in terms of cooling
sweeps). After about 20 cooling sweeps the monopoles become essentially static.
Pairs that have a large separation have no chance of annihilating and further
cooling evolves these configurations to some constant non-zero action. For
QED$_3$ it is possible therefore to classify gauge field configuration
according to whether or not they contain monopoles or antimonopoles that
survive cooling.

The observables calculated are the string tension and static potential obtained
from Wilson loops, magnetic field correlations which are related to the
spin-spin interaction and the chiral symmetry breaking order parameter for
staggered fermions. The results are given in Section III. The behavior of
Wilson loops has some similarity with what was found in QCD. Small Wilson loops
cool much more rapidly than large loops. The Creutz ratios of large loops do
not change much under cooling so the string tension survives with only a small
decrease from the uncooled value. This effect is associated in QED$_3$ with the
fact that plasma monopoles become frozen into the gauge configurations and has
been noted previously by Duncan and Mawhinney\cite{duncan91}.

We can see more directly how monopoles affect the long distance behavior of
QED$_3$ by considering the static potential. Separating our ensemble of gauge
field configurations into a sample that contains monopoles that survive cooling
(plasma monopoles) and a sample in which all monopoles-antimonopoles pairs have
small separations and annihilate, it was found that the static potential for
these two samples behaves quite differently at large distances, {\em even
without any cooling}. Only in the sample with plasma monopoles is there any
evidence of confining behavior.

In contrast to the static potential, magnetic field correlations, which are
related to
the spin-spin interaction\cite{michael,katz}, are essentially the same in
uncooled configuration with or without plasma monopoles. Cooling reveals that
monopoles (recall that these are the instantons of compact QED$_3$) play a
very minor role in the spin-spin interaction. Since it has been suggested that
instantons of QCD are important in the spin-dependent potential of that
theory\cite{shif,shur,oka}, there is a natural application of the methods
used here.

Finally chiral symmetry breaking in (quenched) compact QED$_3$ was examined.
Again by considering configurations with and without plasma monopoles evidence
is presented which suggests that, on the lattice, the presence of widely
separated mono\-pole\--anti\-mono\-pole pairs dominates dynamical mass
generation. A small part of the chiral symmetry breaking comes from bound
monopoles or quantum fluctuations effects.

\section{METHOD}

The usual plaquette action
\begin{eqnarray}
S=\beta \sum_{x, \mu > \nu} \{1-\cos (\theta_{\mu \nu}(x))\}
\end{eqnarray}
is used for compact QED$_3$. The quantity $\beta$ is the dimensionless coupling
constant $\frac{1}{e^2}$ in lattice units and the plaquette angle is
\begin{eqnarray*}
\theta_{\mu \nu}(x) = \theta_\mu(x) + \theta_\nu(x+\hat{\mu}) -
\theta_\mu(x+\hat{\nu}) - \theta_\nu(x)
\end{eqnarray*}
in terms of link variables. Periodic boundary conditions in all directions were
used for the gauge field. Field configurations were constructed using a
heatbath algorithm.

Compact lattice QED contains field configurations with nontrivial topology. In
our 2+1- dimensional Euclidean space these are instantons but we follow the
conventional nomenclature and use the term monopoles since these instantons
 are also the lattice analogs of Dirac monopoles. A convenient
definition of lattice monopoles is that of DeGrand and Toussaint\cite{degrand}.
The plaquette angles are written as
\begin{eqnarray}
\theta_{\mu \nu}(x) = \overline{\theta}_{\mu \nu}(x) + 2\pi \,\, \eta_{\mu
\nu}(x)
\end{eqnarray}
where the reduced plaquette angle $\overline{\theta}_{\mu \nu}(x) \, \in
(-\pi, \pi]$. Summing the oriented reduced plaquette angles associated with
elementary cubes of the lattice allows one to find the monopoles essentially by
identifying those cubes which contain the end of a Dirac string. For a finite
lattice with periodic boundary conditions the number of antimonopoles equals
the number of monopoles. On a finite lattice it can also happen that a monopole
and antimonopole can annihilate leaving a ``magnetic" flux loop that winds
around the lattice\cite{grosch}. To cope with the possible metastability
of such configuration Damgaard and Heller\cite{dang} suggested doing periodic
global Metropolis updates which introduce random flux loops. We use this
procedure here.

Field configurations generated by the heatbath Monte Carlo were cooled using a
local cooling algorithm containing a parameter $\delta$ which controls the rate
of cooling. Cooling reduces the action monotonically by replacing links one at
a time. Consider some particular link, characterized by an angle, call it
$\theta$, which is to be updated. The local contribution to the action can be
written as
\begin{eqnarray}
\tilde{S} = \beta \, (1-r\cos(\theta+\theta_s))
\end{eqnarray}
where $re^{i\theta_s}$ is the sum of the ``staples" of the plaquettes
containing the link being updated. Clearly, $\theta \to \theta^\prime = -
\theta_s$ minimizes $\tilde{S}$. However to control rate of cooling we adopt
the following prescription
\begin{eqnarray}
\global\def\theequation{4a}
\theta \to \theta^\prime &=& \theta + \Delta \theta, \\
\global\def\theequation{4b}
\mid \Delta \theta \mid &=& {\rm min} \{ \delta, \mid \theta + \theta_s \mid
\},\\
\global\def\theequation{4c}
{\rm sign} \Delta \theta &=& -{\rm sign} (\theta + \theta_s) .
\end{eqnarray}
This is the U(1) analog of the algorithm used by Campostrini \emet\cite{campo}.
In the numerical calculations, $\delta$=0.05 was used as in Ref.\cite{campo}.

The results presented in this paper were calculated using cooling sweeps in
which links were updated in a fixed sequence. However some tests were done to
see if cooling links in a random way introduced any differences. It was found
that for a given starting configuration, the final position of a ``frozen in"
monopole or antimonopole might occasionally differ by one lattice site between
sequential and random updating. Also, in rare configurations the monopole
number differs after cooling due to the annihilation of one more or fewer
monopole-antimonopole pairs using fixed versus random updating. However this
happens sufficiently infrequently that we expect ensemble averages over large
samples to be unchanged.

The behavior of a number of observables was monitored under cooling. First, the
Creutz ratio
\begin{eqnarray}
\global\def\theequation{\arabic{equation}}
\setcounter{equation}{5}
C(R,T)=-ln\, \frac{W(R,T)W(R-1,T-1)}{W(R-1,T)W(R,T-1)}\,\,,
\end{eqnarray}
where $W(R,T)$ denotes the $R$ by $T$ rectangular Wilson loop, can be used to
determine the string tension. For large loops, which obey the area law,
 $C(R,R)$
gives the string tension directly. The  potential $V(R)$ between
static charges is also considered. It is calculated by extrapolating Wilson
loops to large $T$
\begin{eqnarray}
V(R)= - {\lim_{T \to \infty}}
\,\, \frac{1}{T} \,\, ln \,\, W(R,T).
\end{eqnarray}

In addition to the confining central interactions Dirac fermions will also have
spin-spin and spin-orbit interactions, even in two spatial dimensions. These
can be calculated on the lattice\cite{michael}. In particular the spin-spin
interaction is related to magnetic field correlations (see Ref.\cite{katz} for
a simple derivation). In two spatial dimensions the magnetic field has only one
component related to F$_{12}$, the field strength tensor in the spatial
directions. The magnetic field correlation is then calculated using a Wilson
loop (say in the $X-T$ plane) with insertion of spatial plaquettes in the
time legs as shown in the Fig. 1. Let $<B(0,t_1) B(R, t_2)>$ denote the
configuration average of the type illustrated in Fig. 1. The correlation
function we calculate is
\begin{eqnarray}
\frac{1}{T-1} \sum_{t_1=T/2, T/2 \pm1} \,\, \sum_{t_2=1, T-1} \,\, \frac{<B(0,
t_1) B(R, t_2)>} {W(R,T)}
\end{eqnarray}
which, up to some ($R$-independent) factors, gives an estimate of the spin-spin
interaction. In practise the magnetic field insertion $B$ that was used was the
average over the four spatial plaquettes whose corners lie on the Wilson loop
$W(R,T)$ corresponding to operator II of Ref.\cite{michael}.

Chiral symmetry breaking was also calculated. Staggered fermions were used with
the usual action
\begin{eqnarray}
\global\def\theequation{8a}
S_f&=&\frac{1}{2} \sum_{x,u} \eta_\mu(x) [\overline{\chi}(x) \, e^{i
\theta_\mu(x)}
\,\chi_{(x+\hat{\mu})} \,\, -\overline{\chi}_{(x+\hat{\mu})}
e^{i\theta_\mu(x)} \, \chi(x)] + \sum_x \, m \overline{\chi}(x) \chi(x), \\
\global\def\theequation{8b}
&=& \mbox{} \overline{\chi} \,\, M(\{\theta\})\chi \, \, .
\end{eqnarray}
where $\overline{\chi}, \chi$ are single component fermion fields,
$\eta_\mu(x)$ is the staggered fermion phase\cite{kawo} and $m$ is the mass in
lattice units. Antiperiodic boundary conditions were used for the fermion
fields in all directions.

The chiral symmetry order parameter is calculated from the inverse of the
fermion matrix $M$ of Eq.(8b)
\begin{eqnarray}
\global\def\theequation{\arabic{equation}}
\setcounter{equation}{9}
<\overline{\chi} \chi>=\frac{1}{V} \, <Tr \, M^{-1}(\{\theta\})>.
\end{eqnarray}
where V is the lattice volume and the angle brackets denote the gauge field
configuration average. A random source method\cite{scalet,fiebig} was used to
calculate $TrM^{-1}(\{\theta\})$. Thirty-two Gaussian random sources were used
for each gauge field configuration.

\section{RESULTS}

The simulation was carried out at $\beta$=2.5 on a $20^3$ lattice. This value
of
$\beta$ was chosen so that a significant fraction of the configurations would
have no monopoles after cooling.

After 5000 heatbath Monte Carlo sweeps to equilibrate, a total of 800
configurations, separated by 150 sweeps, were analyzed. The monopole number
changes relatively slowly and 150 sweeps is roughly the autocorrelation time.
Each configuration was cooled a total of 80 sweeps using the algorithm of
Eq.(4) with $\delta=0.05$. The average plaquette $(<1-\cos \theta_{\mu
\nu}(x)>)$ and the average number of monopoles $N_m$ (recall number of
anti\-mono\-poles = number of monopoles) for the full sample of 800
configurations are plotted in Fig.~2 as a function of cooling sweeps. By about
30 cooling sweeps the average plaquette and monopole number have become
essentially constant.

To better understand what is happening one needs more information about the
monopole excitations in the vacuum. This is provided in Fig. 3. The
monopole-antimonopole correlation parameter $C_m$ counts the number of times a
monopole, located in an elementary cube on the lattice, has an antimonopole in
a neighboring cube. From Fig. 3a it is seen that all ``bound"
mono\-pole\--anti\-mono\-pole pairs, i.e., pairs on neighboring cubes,
annihilate under
cooling. Comparing $C_m$ and $N_m$ one sees that only a small fraction of the
monopoles that are not initially in neighboring cubes annihilate. A more
detailed examination of a few configurations showed that pairs separated by a
distance greater than $\sqrt{3}$ lattice units, i.e., without a face, edge or
point in common, very rarely annihilate. This reflects the lack of mobility of
the monopoles under slow local cooling. After cooling what we call plasma
monopoles are left. This is shown in another way in Fig. 3b where the average
minimum separation between a monopole and the nearest antimonopole is plotted.

Figure 4 shows the Creutz ratio $C(R,R)$ for the full set of configurations
with
no cooling and with 16, 32 and 80 cooling sweeps. The features are the same as
observed in previous calculations\cite{campo,duncan91}. Small Wilson loops
cool much more rapidly than large loops. The statistical fluctuations also
decrease very rapidly making the string tension visible after a small amount of
cooling. However, as also noted by Duncan and Mawhinney\cite{duncan91}, the
apparent string tension from slightly cooled configurations may be an
overestimate of the true asymptotic string tension.

We would like to see the role of monopoles as directly as possible. Since at
least some monopoles survive cooling it is natural to classify configurations
according to whether they contain monopoles and antimonopoles after cooling
$(N^{c}_{m} \neq 0)$ or whether they do not $(N^{c}_{m} = 0)$. Recall
that configurations in which monopole-antimonopole pairs annihilate leaving
flux wrapped around the lattice can occur. These are excluded from the
$N^{c}_{m}=0$ sample. Out of a total 800 configurations, 488 went into the
$N^{c}_{m} \neq 0$ sample and 145 had $N^{c}_{m} = 0$.

The properties of the two samples, with and without monopoles that survive
cooling, are compared in Fig. 5 and 6. Note in particular Fig. 6b which shows
for the $N^{c}_{m} = 0$ sample that it is essentially only
monopole-antimonopole pairs separated by less than two lattice units that
annihilate.

The Creutz ratio $C(R,R)$ for the $N^{c}_{m} =0$ and $N^{c}_{m} \neq 0$
ensembles are compared in Fig. 7 using uncooled configurations. Results
with 16, 32 and 80 cooling sweeps are shown in Fig. 8. The large Wilson loops
with no cooling are very noisy so a definitive statement is not possible but a
trend is evident in Fig. 7: configurations without widely separated (plasma)
monopole-antimonopole pairs show no sign of a string tension. With a small
amount of cooling the situation becomes quite clear: without monopoles (Fig.
8b) large Wilson loops are not suppressed by an area law, i.e., the Creutz
ratio becomes trivial. With monopoles (Fig. 8a), Wilson loops on the scale of
the average monopole-antimonopole separation are suppressed and a string
tension remains even with extreme cooling.
Evidently with extreme cooling the string tension can be extracted
from $C(R,R)$ only for $R$ much larger than the average
monopole-antimonopole separation.

The above effect is seen even more nicely by considering the static potential
directly. Figure 9 shows the static potential, Eq.(6), calculated using
$N^{c}_{m} \neq 0$, $N^{c}_{m} = 0$ and full uncooled gauge configuration
samples. At short distances all potentials are the same suggesting that quantum
fluctuations dominate\cite{faber}. However the $N^{c}_{m} = 0$ sample, from
which plasma monopoles are excluded, yields a potential which shows a
pronounced flattening at large distance. After cooling the situation is shown
in Fig. 10. The string tension extracted from the last few points of the
uncooled $N^{c}_{m} \neq 0$ potential (open squares) is about 0.018 while
after 80 cooling steps (filled squares) a string tension of about 0.013 is
obtained. The Creutz ratio $C(R,R)$ (Fig. 8a) is
apparently just in between which is
reasonable since the potentials with cooled and uncooled configurations
approach
the limiting linear behavior with the opposite curvature\cite{wensley}.

In QCD we normally consider the spin-spin interaction as due to the exchange of
gluons. However there exist suggestions that a substantial part of the
hyperfine interaction is due to instanton effects\cite{shif,shur,oka}.
If this is so, we would expect magnetic field correlations, which measure the
spin-spin interaction\cite{katz}, to be very different in configurations with
different instanton properties. Also strong magnetic field correlations
would be expected to persist if instantons survive cooling. A calculation
to test this directly has not yet been done for QCD (see however
Ref.\cite{chu}). For QED$_3$ the results are shown in Fig. 11 and 12. The
squares in Fig. 11 show the magnetic field correlation (Eq.(7)) calculated
using
the uncooled $N^{c}_{m} \neq 0$ gauge field sample. The $N^{c}_{m} = 0$
sample was further divided into two sets of configurations: those which contain
monopoles which annihilate under cooling and those configurations which contain
no monopoles even before cooling. The magnetic field correlation calculated
with these two sets of uncooled configurations is shown in Fig. 11 by triangles
and circles respectively. Essentially no difference is discernible between the
three calculations. Note also the magnetic field correlation observed here is
qualitatively the same as seen in QCD$_4$ (compare, for example, with the
``operator II" results of Fig. 3 in Ref.\cite{michael}). Figure 12 shows the
magnetic field correlation for the $N^{c}_{m} \neq 0 $ sample after 80
cooling sweeps (note the change in scale from Fig. 11). For QED$_3$ at least,
monopoles (instantons) apparently play a very small role in the spin-spin
interaction which is due predominantly to short distance quantum fluctuations.

In addition to linear confinement of charge, compact QED$_3$ has another
property which is also important in QCD$_4$, namely, chiral symmetry breaking.
It is natural to seek a common mechanism behind these two
phenomena\cite{casher}. We calculated the chiral symmetry breaking parameter
$<\overline{\chi}\chi>$ for staggered fermions for a subset (400) of our
quenched gauge field configurations. The mass range for the calculations was
0.02 to 0.1 in lattice units. A nonzero value in the limit of zero fermion
mass indicates chiral symmetry breaking. The results for uncooled
configurations are shown in Fig. 13.
As with the magnetic field correlation it is instructive to consider three
subsets of configurations: configurations containing monopoles after cooling
($N^c_m \neq 0)$, configurations containing only monopoles that annihilate
when cooled and those  with no monopoles at all.  Figure 13 shows that the
presence of plasma monopoles, which leads to linear confinement, also
significantly enhances dynamical mass generation.  Configurations containing
mono\-pole\--antimonopole pairs separated only by short distances give
essentially the same value of $<\overline{\chi}\chi>$ as configuration that
contain only nontopological quantum fluctuations.  In the continuum limit,
$\beta \to \infty$, monopole fluctuations disappear\cite{degrand} but chiral
symmetry breaking is expected to persist (see \cite{curtis} and references
therein).  Our results are qualitatively consistent with this expectation.
Determining a precise value for $<\overline{\chi}\chi>$ at zero fermion mass is
very difficult (see example Ref. \cite{hands}) but a qualitative conclusion,
for example, by linearly extrapolating from the last few mass points in Fig.
13, is that $<\overline{\chi}\chi>\!|_{m=0}$ is nonzero in all subsets of
configurations, i.e., nonopological quantum fluctuation can give
chiral symmetry breaking. At $\beta = 2.5$ dynamical mass generation however
seems to be dominated by plasma monopoles even though the density of such
 monopoles is not very large.  After 80 cooling surveys we get the results in
Fig. 14.  Chiral symmetry breaking persists in those configurations which
contain plasma monopoles not annihilated under cooling.

\section{SUMMARY}

In this paper we show how cooling can be used to correlate the long-distance
confining behavior of compact QED$_3$ with the presence of separated
monopole-antimonopole pairs in the vacuum. Such plasma monopoles survive
cooling in QED$_3$ and therefore it is possible to use their presence as a way
of classifying gauge field configurations. The difference between our work and
previous studies is that after determining the monopole properties by cooling
we go back to recalculate and compare observables in subsets of gauge field
configuration with different long-distance monopole characteristics. What
emerges is a consistent picture: the Creutz ratio $C(R,R)$ and the static
potential $V(R)$ are determined at large $R$ by the presence of
monopole-antimonopole pairs with separation comparable to $R$. Cooling out
quantum fluctuations or monopole-antimonopole pairs with small separation does
not change the long distance behavior appreciably.

Chiral symmetry breaking, which is also a long-distance phenomenon, was also
seen to be dominated at $\beta = 2.5$ by plasma monopoles. On the other hand,
magnetic field correlations which determine the spin-spin interaction seem to
be determined by short-distance effects. After cooling, only a very small
correlation from plasma monopoles remains.

Our study of QED$_3$ points to some calculations that can be done in QCD$_4$.
Obviously instanton effects can be looked for in the same way as
done here. However there are other kinds of topological excitations that one
may try to examine, for example Abelian monopoles\cite{abelian} or
vortices\cite{vort}. These type of objects apparently do not get frozen in
under cooling\cite{deld} so some modification of our method would be necessary.
However one lesson can be learned from the present study, that is, one needs to
correlate the behavior of observables with more detailed vacuum properties
rather than simply focussing on, for example, the monopole density as has been
done in many papers up to now\cite{deld,density}.

\section{ACKNOWLEDGEMENTS}

One of us (R.M.W.) thanks T. Fujita and M. Oka for some helpful discussion.
This work was supported in part by the Natural Sciences and Engineering
Research Council of Canada.

\newpage


\bibliographystyle{unsrt}      

\newpage

\begin{center}
{\bf Figure Captions}
\end{center}

\begin{enumerate}
\item Example of a Wilson loop with spatial plaquette insertions.

\item (a) average plaquette and (b) number of monopoles $N_m$ as a function of
cooling sweep for the full sample of 800 configurations.

\item (a) Number of neighboring monopole-antimonopole pairs $C_m$ and (b)
average minimum monople-antimonopole separation $<r_{\rm min}>$ as a function
of cooling sweep for the full sample of 800 configurations.

\item The Creutz $C(R,R)$ as a function of loop size $R$ for the full
sample of 800 configurations with no cooling ($\triangle$),
16 cooling sweeps ($\triangle$), 32 cooling sweeps ($\bullet$) and 80 cooling
sweeps (\rule{2mm}{2mm}).

\item (a) average plaquette and (b) number of monopoles $N_m$ as a function of
cooling sweep for the $N^c_m \neq 0$ sample of 488 configurations ($\Box$) and
the $N^c_m = 0$ sample of 145 configurations ($\circ$).

\item (a) number of neighboring monopole-antimonopole pairs $C_m$ and (b)
average minimum monopole-antimonopole separation $<r_{\rm min}>$ as a  function
of cooling sweep for the $N^c_m \neq 0$ sample of 488 configurations ($\Box$)
and the $N^c_m = 0$ samples of 145 configurations ($\circ$).

\item The Creutz ratio $C(R,R)$ as a function of loop size $R$ with no
cooling for the $N^c_m \neq 0$ sample of 488 configurations ($\Box$) and the
$N^c_m = 0$ sample of 145 configurations $(\circ)$.

\item The Creutz ratio $C(R,R)$ as a function of loop size with 16 cooling
sweeps $(\triangle)$, 32 cooling sweeps $(\bullet)$ and 80 cooling sweeps
(\rule{2mm}{2mm}) (a) for the $N^c_m \neq 0$ sample and (b) for the $N^c_m =0$
sample.

\item The potential between static charges $V(R)$ versus separation $R$ with no
cooling for the full sample of 800 configurations $(\triangle)$, the $N^c_m
\neq 0$ sample $(\Box)$ and the $N^c_m = 0$ sample $(\circ)$.

\item The potential between static charges $V(R)$ versus separation $R$ for the
$N^c_m \neq 0$ sample with no cooling $(\Box)$ and after 80 cooling sweeps
(\rule{2mm}{2mm}).  Also shown is the result for the $N^c_m = 0$ sample after
80 cooling sweeps $(\bullet)$.

\item The magnetic field correlation (7) versus separation $R$ with no cooling
for the $N^c_m =0$ sample $(\Box)$, a subset of the $N^c_m =0$ configuration of
the configurations which contain monopoles that annihilate when cooled
$(\circ)$ and a subset of the $N^c_m = 0$ configurations which contain no
monoples even before cooling $(\triangle)$.

\item The magnetic field correlation (7) versus separation $R$ after 80 cooling
sweeps for the $N^c_m \neq 0$ sample.

\item The chiral order parameter $<\overline{\chi}\chi>$ versus fermion mass
$m$ (in lattice units) with no cooling calculated for a subset of $N^c_m \neq
0$ configurations $(\Box)$ a subset of $N^c_m=0$ configurations which contain
monopoles that annihilate when cooled ($\circ$) and subset of $N^c_m =0$
configurations which contain no monoples even before cooling $(\triangle)$.

\item  The chiral order parameter $<\overline{\chi}\chi>$ versus fermion mass
$m$ (in lattice units) after 80 cooling sweeps calculated for a subset of
$N^c_m \neq 0$ configurations (\rule{2mm}{2mm}), a subset of $N^c_m =0$
configuration which contain monopoles that annihilate when cooled ($\bullet$)
and a subset of $N^c_m =0$ configurations which contain no monopoles even
before cooling $(\triangle)$.

\end{enumerate}

\end{document}